\documentclass[pra,twocolumn,aps,letterpaper, preprintnumbers,superscriptaddress,amsmath,amssymb]{revtex4}
\usepackage{graphicx}
\usepackage{bm}
\usepackage{siunitx}
\usepackage{xcolor}
\usepackage{ulem}
\usepackage{hyperref}
\usepackage{color}

\begin{document}
\title{Tweezing microparticles by tightly focused optical skyrmionic fields}
\author{Liuhao Zhu}
 \affiliation{School of Electronic Science and Engineering, Southeast University, Nanjing 211189, China}
 \affiliation{Centre for Disruptive Photonic Technologies, School of Physical and Mathematical Sciences, Nanyang Technological University, Singapore 637371, Singapore}
\author{Hao Zhang}
\affiliation{Centre for Disruptive Photonic Technologies, School of Physical and Mathematical Sciences, Nanyang Technological University, Singapore 637371, Singapore}
\author{Xi Xie}
\email{xiexi@cdut.edu.cn}
\affiliation{College of Physics, Chengdu University of Technology, Chengdu, 610059, China}
\affiliation{Centre for Disruptive Photonic Technologies, School of Physical and Mathematical Sciences, Nanyang Technological University, Singapore 637371, Singapore}
\author{Yanhe Lyu}
\affiliation{Centre for Disruptive Photonic Technologies, School of Physical and Mathematical Sciences, Nanyang Technological University, Singapore 637371, Singapore}
\author{Bing Gu}
\email{gubing@seu.edu.cn}
\affiliation{School of Electronic Science and Engineering, Southeast University, Nanjing 211189, China}
\author{Etienne Brasselet}
\email{etienne.brasselet@u-bordeaux.fr}
\affiliation{CNRS, Laboratoire Ondes et Matière d’Aquitaine, University of Bordeaux, 33400 Talence, France}
\author{Yijie Shen}
\email{yijie.shen@ntu.edu.sg}
\affiliation{Centre for Disruptive Photonic Technologies, School of Physical and Mathematical Sciences, Nanyang Technological University, Singapore 637371, Singapore}
\affiliation{School of Electrical and Electronic Engineering, Nanyang Technological University, Singapore 639798, Singapore}

\date{\today}

\begin{abstract}
The concept of skyrmions, once translated to the optical domain, provides a topological framework to investigate structured light-matter interactions. However, their practical use for the manipulation of material systems remains barely explored. Here, we experimentally demonstrate trapping and tunable orbital motion of dielectric microparticles in tightly focused skyrmionic polarization textures prepared using a compact spin-orbit optical element. We explore the tweezing capabilities at fixed topology either by varying the relative weights of the orthogonally polarized components or by switching between skyrmion and bimeron polarization textures. Vectorial focusing and optical-force calculations qualitatively reproduce the experimental observations. These results motivate extending skyrmionic light to the manipulation of micro-objects, including birefringent and chiral samples.
\end{abstract}

\maketitle

\noindent

\section{Introduction}

Optical tweezers provide a versatile platform for the noncontact trapping and manipulation of microscopic particles through light-matter linear and angular momentum transfer \cite{ashkinObservationSinglebeamGradient1986, volpeRoadmapOpticalTweezers2023}. Beyond conventional free-space optical tweezers, plasmonic platforms enable nanoscale optical manipulation \cite{zhangPlasmonicTweezersNanoscale2021}, while hollow-core optical fibers extend optical trapping and transport over long distances \cite{wangOpticalTrappingMesoscale2025}. Advances in holographic wavefront shaping and structured-light engineering have expanded control over spatial phase, polarization, and optical angular momentum \cite{yangOpticalTrappingStructured2021, otteOpticalTrappingGets2020, padgettTweezersTwist2011, maObservationSuperoscillationMoire2026}, enabling reconfigurable particle arrays \cite{curtisDynamicHolographicOptical2002}, particle rotation \cite{zhuOpticalSkippingRope2023}, orbital transport \cite{gaoParticleDeliveryGeneralized2024a, shenOpticalVortices302019}, curved three-dimensional transport \cite{rodrigoFreestyle3DLaser2015}, sorting \cite{yangOpticalSortingPresent2025}, and multidimensional manipulation \cite{zhuMultidimensionalOpticalTweezers2023}. Vector-field engineering further enables rotation about arbitrary three-dimensional axes and orbital transport driven by polarization gradients \cite{wuTimeVarying3DOptical2025, yuanSpinEngineered3DOptical2025}. Longitudinal orbital-angular-momentum engineering and energy-streamline design extend control over optical momentum and particle motion \cite{yuanLongitudinallyTunedOAM2026, yanStructuringLightFlows2026}. Meanwhile, holographic plasmonic fields and multifunctional metasurfaces support programmable particle transport and parallel manipulation \cite{zhangStructurelessExcitationManipulation2026, liFreeformOpticalFlow2026, liScalableGeneralizedMetaSpanners2026}. 

Among these structured-light approaches, optical vortices, whose topology lies in the phase of a scalar field, have proved particularly powerful for trapping and manipulating particles~\cite{padgettTweezersTwist2011}. Stokes skyrmions, which map the transverse plane onto the space of polarization states~\cite{gaoTopologicalLightField2025, shenFreespaceTopologicalOptical2025}, extend topology from the phase of scalar fields to the polarization of vectorial fields. They belong to the broader family of optical skyrmions, which are real-space topological textures of optical vector fields such as the electric and magnetic fields, spin, or energy flow \cite{gaoParaxialSkyrmionicBeams2020, shenGenerationOpticalSkyrmions2022, tsessesOpticalSkyrmionLattice2018, dengObservationLocalizedMagnetic2022, duDeepsubwavelengthFeaturesPhotonic2019, wuPhotonicTorons3D2025, chenGouyPhaseInduced2025, shenFreespaceTopologicalOptical2025}. A prototypical Stokes skyrmion is the coherent superposition of two orthogonally polarized spatial modes, a Gaussian mode and a Laguerre-Gauss vortex mode \cite{shenGenerationOpticalSkyrmions2022}. The generation \cite{tsessesOpticalSkyrmionLattice2018, dengObservationLocalizedMagnetic2022}, propagation \cite{guoSelfhealingOpticalSkyrmionic2025}, and tight-focusing properties \cite{duDeepsubwavelengthFeaturesPhotonic2019, wuPhotonicTorons3D2025,chenGouyPhaseInduced2025} of such textures have been widely studied, and spin-orbit optical elements now provide compact ways to produce them \cite{hakobyanPlatesOpticalVortices2025, hakobyanUnitarySpinorbitOpticalskyrmionic2024}. These advances make skyrmionic fields attractive for controlling and studying light–matter interactions \cite{xieTopologicalLightWaves2025, luNanoparticleDeepSubwavelengthDynamics2024}.

Because optical forces and torques follow the spatial structure of the field, skyrmionic light fields are natural candidates for optical manipulation. Theory and simulations predict deep-subwavelength confinement, size-selective transport and sorting, and reconfigurable trapping~\cite{luNanoparticleDeepSubwavelengthDynamics2024,tangAchiralChiralOptical2023,xuTopologicalMagneticLattices2025,odonnellOpticalForcesTorques2024,hanDynamicCharacteristicsSkyrmionic2026}. Experiments in other wave systems support this picture: topological textures can drive orbital motion and self-rotation in water waves~\cite{smirnovaWaterWaveVorticesSkyrmions2024a,wangTopologicalWaterwaveStructures2025}, co- and counter-rotation in quasicrystalline wave fields~\cite{wangTopologicalWaterwaveStructures2025,duanQuasicrystalTopologicalHydrodynamics2026}, and microparticle trapping and assembly in acoustic skyrmion lattices~\cite{zhangSkyrmionMicroparticleSAW2026}. In optics, however, it remains to be established experimentally how the modal composition of a skyrmion built from mode superposition shapes the focused field and the resulting particle dynamics.

\begin{figure}[t!]
\includegraphics[width=1\columnwidth]{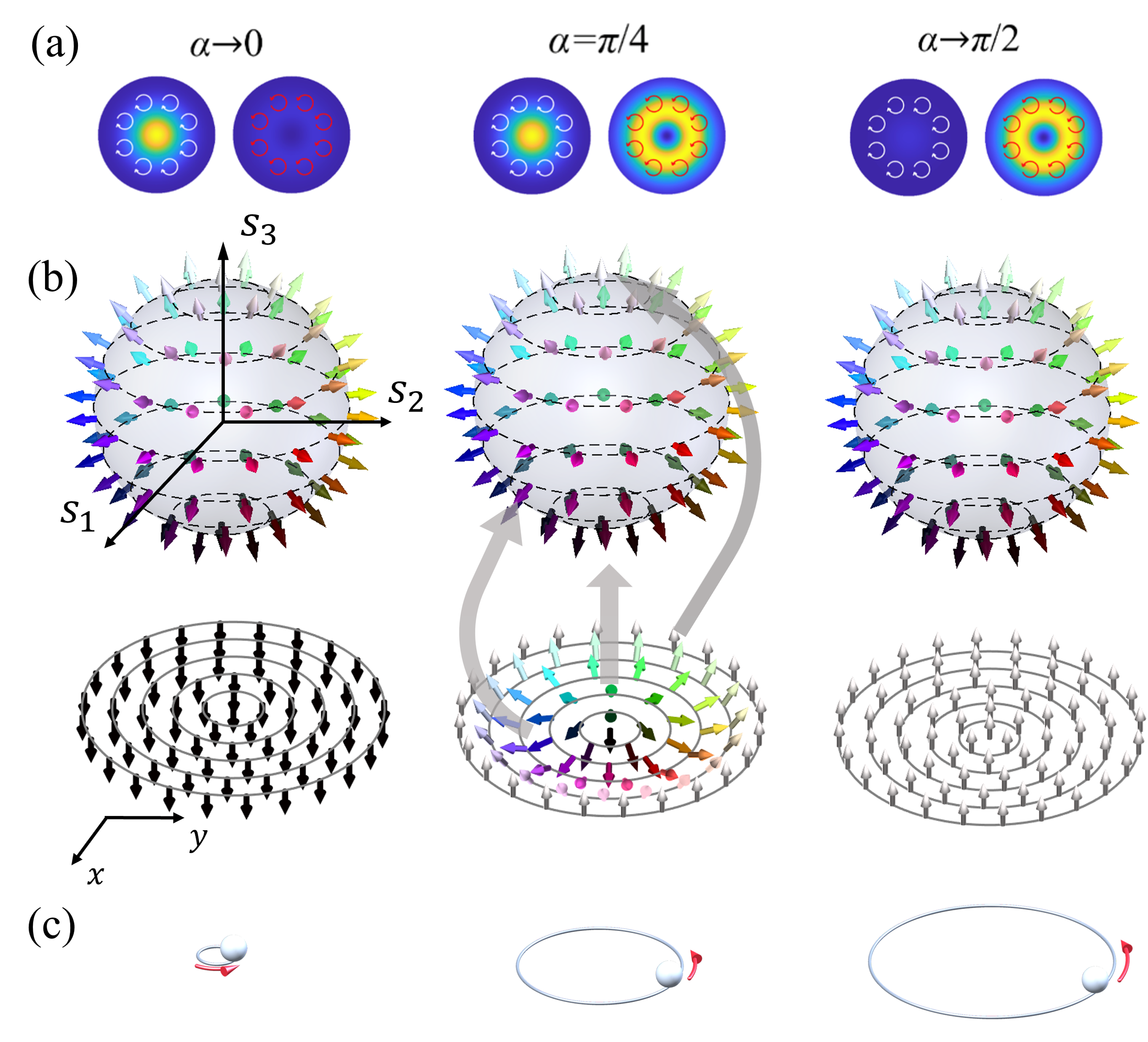}
\caption{\label{fig:Figure1} Optical manipulation of dielectric microparticles in tunable optical skyrmionic fields. A coherent superposition of a right-handed circularly polarized Gaussian mode and a left-handed circularly polarized vortex mode with adjustable non-vortex/vortex modal weight ratio described by the angle $\alpha$ (see text for details) is used as the driving optical field (a). This forms a skyrmionic polarized field with adjustable texture but overall given topology illustrated by mappings between the Poincaré sphere and the spatial distribution of the Stokes vector in the $(x,y)$ transverse plane (b). This allows changing the spatial distribution of the optical force field, which results in controlling the trajectory of the optically trapped microparticles (c).}
\end{figure}

Here, as a first step, we experimentally demonstrate the trapping and tunable orbital motion of dielectric microparticles using tightly focused Stokes skyrmion beams, generated with a compact spin–orbit optical element. We vary the modal composition at fixed topology, through the relative weights of the nonvortex and vortex constituents, as sketched in Fig.~\ref{fig:Figure1}. This reshapes the vector field and modifies the particle motion. We further compare skyrmion and bimeron polarization textures. Vectorial focusing and optical-force calculations qualitatively account for the observations in terms of field redistribution, confinement, and azimuthal drive.

\section{Results}
\label{sec:results}

\subsection{Preparation of collimated adjustable Stokes skyrmionic fields}

\begin{figure}[t!]
\includegraphics[width=\columnwidth]{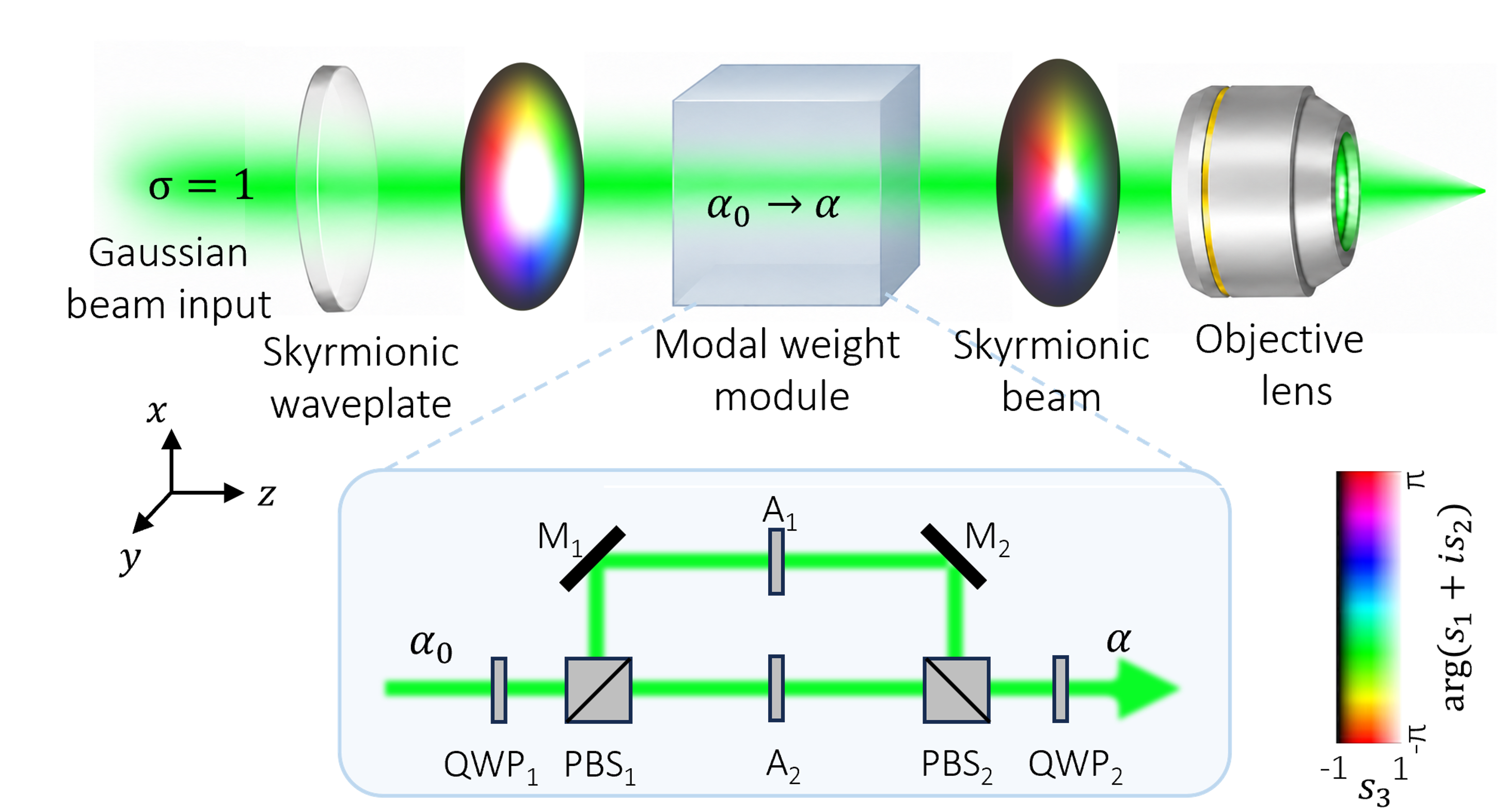}
\caption{\label{fig:Figure2} Sketch of the generation, manipulation, and focusing of the optical skyrmionic field from a circularly polarized Gaussian beam. The modal weight module controls the optical power ratio between the NV and V components of the skyrmionic field $P_{\rm V}/P_{\rm NV}=\tan^2\alpha$. The texture sketches corresponds to $\sigma=+1$.}
\end{figure}

Experimentally, the skyrmionic field is prepared following the scheme illustrated in Fig.~\ref{fig:Figure2}(a), using a first-order unitary spin-orbit optical skyrmionic wave-plate introduced in Ref.~\cite{hakobyanUnitarySpinorbitOpticalskyrmionic2024}. This doubly inhomogeneous wave-plate made from laser nanostructured silica glass is characterized by an optical-axis orientation angle $\psi(\phi)=\phi/2$ and a radially varying birefringent phase retardation $\Delta(\rho)=2\arcsin\left[(\sqrt{2}\rho/w_{\rm design})\exp(1/2-\rho^2/w_{\rm design}^2)\right]$, where $\rho$ and $\phi$ are the radial and azimuthal coordinates, respectively, and $w_{\rm design}=1.5$~mm denotes the Gaussian beam waist radius for which the structured wave plate was designed, for an operating wavelength in vacuum $\lambda_0=532$~nm. The optical element is illuminated at normal incidence by a circularly polarized fundamental Gaussian beam that corresponds to the input field $\mathbf E_{\mathrm{in}}=E_0\exp(-\rho^2/w_{\rm in}^2)\,\mathbf e_{\sigma}$, where $\mathbf e_{\sigma}=({\bf e}_x + i\sigma {\bf e}_y)/\sqrt{2}$ denotes the unit circular polarization vector with helicity $\sigma=\pm 1$, the field emerging from the used unitary skyrmionic wave-plate is 
\begin{eqnarray} 
\nonumber
&&\mathbf E_{\mathrm{out}}(\rho,\phi)
= E_0 \exp[-\rho^2/w_{\rm in}^2+ i\Phi_{\rm dyn}(\rho)]\\
&&\times\left\{
\cos\left[\Delta(\rho)/2\right]\,\mathbf e_{\sigma} +i\sin\left[\Delta(\rho)/2\right]
 e^{i\sigma \phi}\,\mathbf e_{-\sigma}\right\},
\label{eq:plate_output}
\end{eqnarray}
where $\Phi_{\rm dyn}(\rho) = -0.65\Delta(\rho)$ is the radially dependent dynamic phase arising from the  retardance radial profile, which is specific to the used fabrication method \cite{hakobyanUnitarySpinorbitOpticalskyrmionic2024}. Even if not implemented in the present work, dynamic phase compensation strategies, either spin-dependent \cite{hakobyan2024single} or spin-independent \cite{chavilkkadanPurelyGeometricSpin2025} have been recently developed. Then, the relative weight of the orthogonally polarized non-vortex (NV) and vortex (V) components of the generated skyrmionic field is controlled by a modal weight module described in Fig.~\ref{fig:Figure2}. The latter consists of a quarter-wave plate (QWP) followed by a polarizing beam splitter (PBS), which first separates the NV and V components into two distinct optical paths. The relative modal weights are then adjusted using a variable optical attenuator (A) placed in each arm. Finally, the two components are recombined by a second PBS-QWP set, producing the modified skyrmionic field.

\begin{figure}[b!]
\includegraphics[width=\columnwidth]{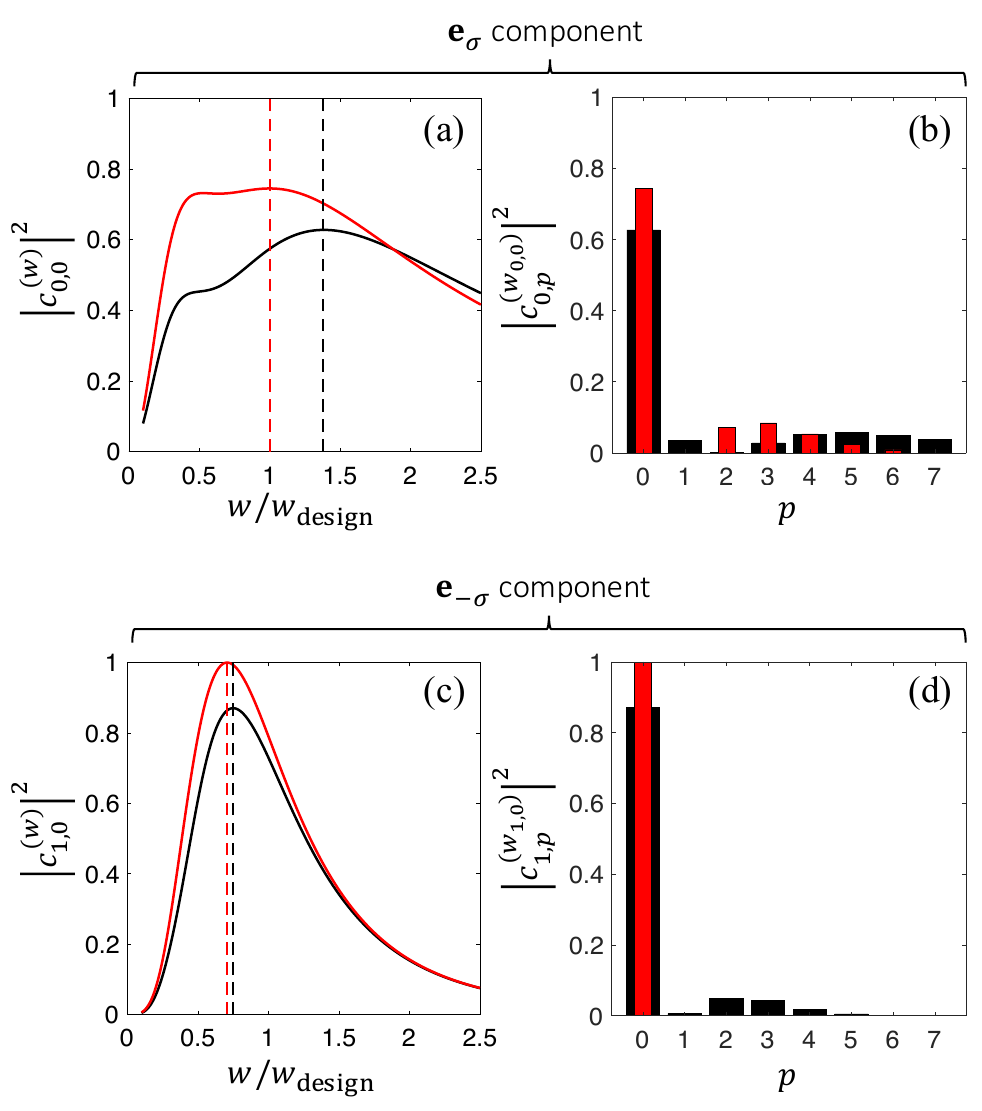}
\caption{\label{fig:Fig_modality} Laguerre-Gauss modal decomposition of the $\sigma$-circularly polarized component of the optical field before being focused by the microscope objective for the ideal (red color) and experimental (black color) situations that respectively correspond to $w_{\rm in} = w_{\rm design}$ and $w_{\rm in} = 1.17w_{\rm design}$. (a) Calculated modal power fraction of the fundamental Gaussian mode as a function of the normalized waist radius chosen for the decomposition. The dashed lines denote the optimal waist $w_{0,0}$ maximizing the Gaussian modal power fraction. (b) Modal decomposition over the radial modes ${\rm LG}_{0,p}^{(w_{0,0})}$ for $w_{\rm in}=w_{0,0}$, where $w_{0,0\,\rm ideal}=w_{\rm design}$ and $w_{0,0\,\rm exp}=1.38w_{0,0\,\rm ideal}$. (c,d) Same as panels (a,b) for the $-\sigma$-circularly polarized component giving $w_{1,0\,\rm ideal}=w_{\rm design}/\sqrt{2}$ and $w_{1,0\,\rm exp}=1.06w_{1,0\,\rm ideal}$.}
\end{figure}

A physically insightful approximation of the field incident on the focusing microscope objective, given by Eq.~(\ref{eq:plate_output}), is obtained by retaining only the dominant spatial modes of the NV and V components. Within the paraxial approximation, the field is expanded onto the orthonormal Laguerre-Gaussian basis, whose modes, denoted by ${\rm LG}_{l,p}^{(w)}$, are characterized by an azimuthal index $l \in \mathbb{Z}$, a radial index $p \in \mathbb{N}$, and a Gaussian beam waist radius $w$ (see Appendix A). The results are shown in Fig.~\ref{fig:Fig_modality} for both circular components, where $|c_{l,p}^{(w)}|^2$ denotes the power fraction of mode ${\rm LG}_{l,p}^{(w)}$ in the corresponding circular component. We also define $w_{0,0}$ as the waist radius that maximizes the contribution of mode $(l,p)=(0,0)$ for the component ${\bf e}_\sigma$, and $w_{1,0}$ as the waist radius that maximizes the contribution of mode $(l,p)=(1,0)$ for the component ${\bf e}_{-\sigma}$. As shown in the figure, the ideal case, assuming a flat dynamic phase profile and an incident beam waist $w_{\rm in}=w_{\rm design}$ (red color), corresponds to a dominant-mode content very similar to that obtained for the experimental implementation (black color), which uses $w_{\rm in}=1.17w_{\rm design}$ and includes the effect of uncompensated dynamic phase modulation. This indicates that the ideal model provides a good approximation of the experimental behavior while also being directly applicable to future skyrmionic wave plates with a flat dynamic phase profile. Therefore, in the following, we retain the ideal dominant-mode expression for the field incident on the microscope objective,
\begin{equation}
{\bf E}=\cos\alpha\,\,{\rm LG}_{0,0}^{(w_{\rm in})}\,{\bf e}_{\sigma}
+ i \sin\alpha\,\,{\rm LG}_{1,0}^{(w_{\rm in}/\sqrt{2})}\,{\bf e}_{-\sigma}\,,
\label{eq:modal-weight}
\end{equation}
where the angle $\alpha$ parametrizes the power fraction of the two modes according to $P_{\rm NV}/P_{\rm V}=\tan^2\alpha$. Before entering the modal weight module, one gets from Eq.~(\ref{eq:plate_output}) with $w_{\rm in}=1.17w_{\rm design}$ that $\alpha_0 \simeq 0.30\pi$, whereas we experimentally measure $\alpha_0 \simeq 0.26\pi$. From this initial value, any value of $\alpha$ between $0$ and $\pi/2$ can be achieved by tuning the attenuators A$_1$ and A$_2$ before being sent to the microscope objective (see Fig.~\ref{fig:Figure2}).

\begin{figure*}[t!]
\includegraphics[width=2\columnwidth]{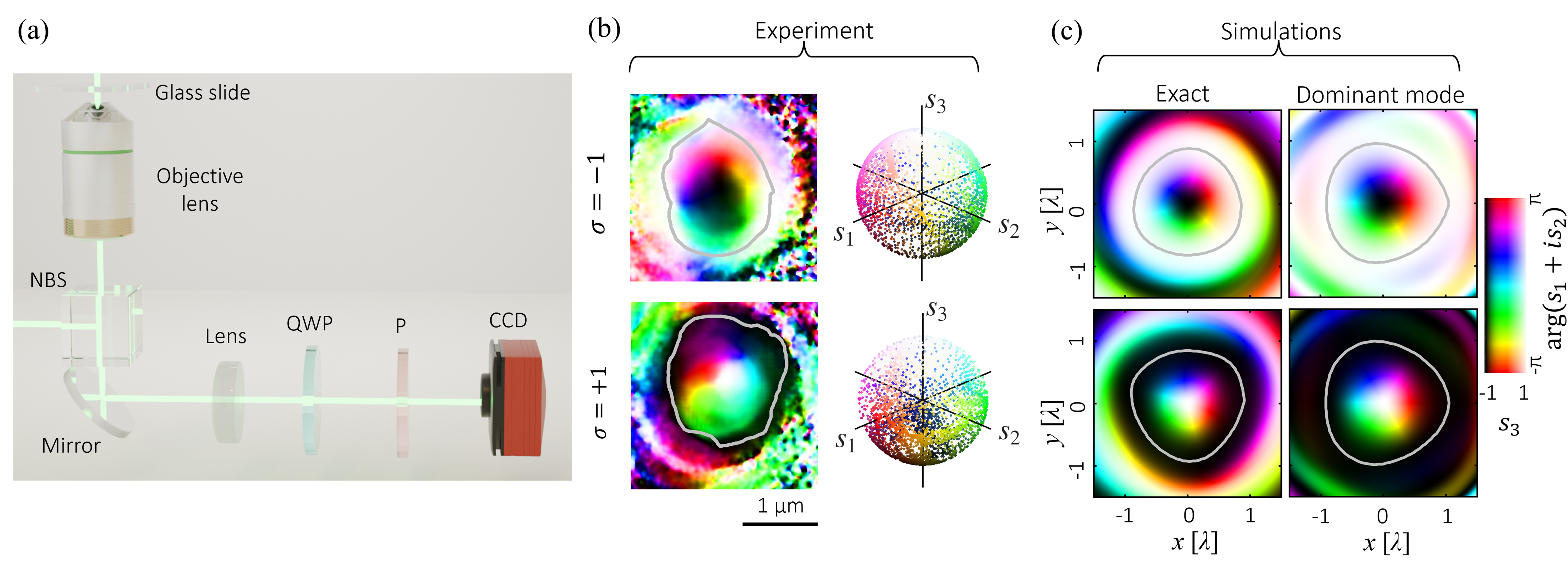}
\caption{\label{fig:Figure4} (a) Experimental setup sketch used to measure the skyrmionic polarization patterns for an incident circularly polarized beam with helicity $\sigma=\pm1$ on the skyrmionic wave-plate. NBS: non-polarizing beam splitter; P: polarizer. (b) Experimental Stokes textures $\{ {\bf s}\}$ and in the focal plane of the microscope objective and the corresponding coverage of the Poincaré sphere. (c) Simulated Stokes textures $\{ {\bf s}\}$ either from Eq.~\ref{eq:plate_output} (exact field) or Eq.~\ref{eq:modal-weight} with $\alpha = \alpha_0=0.30\pi$ (dominant mode description). The selected areas shown in (b) and (c) correspond to a skyrmion number $N \approx \sigma$.}
\end{figure*}

\subsection{Generation and characterization of tightly focused skyrmionic fields}

Once the field is prepared according to the above procedure, it is focused by a microscope objective Olympus UPlanXApo 100$\times$, ${\rm NA}=1.45$, having an input pupil radius $R_p=2.61$~mm. How well the skyrmionic polarization texture is preserved after focusing is experimentally assessed from the polarimetric analysis of the field back-reflected at the air/glass interface of a glass substrate, as depicted in Fig.~\ref{fig:Figure4}(a) taking care of using a beam splitter that preserves the polarization state. The axial position of the glass slide is manually adjusted using a translation stage while monitoring the reflected beam spot on the camera, whose smallest size indicates that the reflecting surface is aligned with the focal plane of the objective. Spatially resolved Stokes parameters are then measured from intensity patterns $I_i$, $i=({\rm H},{\rm V},{\rm D},{\rm A},{\rm L},{\rm R})$, resulting from sequential projective measurements onto the polarization states defined by the unit vectors ${\bf e}_{\rm H,V}={\bf e}_{x,y}$, ${\bf e}_{\rm D,A}=({\bf e}_{x} \pm {\bf e}_{y})/\sqrt{2}$, and ${\bf e}_{\rm L,R}=({\bf e}_{x} \pm i{\bf e}_{y})/\sqrt{2}$, using a QWP and a polarizer. This gives access to the reduced Stokes vector field ${\bf s}$ defined as
\begin{equation}
{\bf s} = \left(\frac{I_{\rm H}-I_{\rm V}}{I_{\rm H}+I_{\rm V}}, \frac{I_{\rm D}-I_{\rm A}}{I_{\rm D}+I_{\rm A}}, \frac{I_{\rm L}-I_{\rm R}}{I_{\rm L}+I_{\rm R}}\right)\,,
\label{eq:stokes-vector}
\end{equation}
which is subsequently renormalized to compensate for small deviations from unit norm caused by instrumental imperfections. The experimentally reconstructed Stokes textures in the $(x,y)$ plane and on the Poincaré sphere are presented in Fig.~\ref{fig:Figure4}(b) for $\sigma=\pm1$. The corresponding textures calculated from Eqs.~(\ref{eq:plate_output}) and (\ref{eq:modal-weight}) are shown in Fig.~\ref{fig:Figure4}(c). We attribute the angular offset between the experimental and calculated polarization state distributions to a relative phase shift between the two circular components at the output of the modal weight module, induced by the unequal arm lengths. This phase shift rotates the major axis of the polarization ellipse at each spatial point. Since the corresponding path-length difference is negligible compared to the temporal coherence length of the laser source, the field remains fully polarized. The corresponding simulations are carried out using the vectorial Richards-Wolf integral \cite{chenGouyPhaseInduced2025}
\begin{eqnarray}
\nonumber
\mathbf{E}_{\rm focal\,plane}(\rho,\varphi,z) = C\int_0^{\theta_{\max}}\!\int_0^{2\pi}
\mathbf{E}_{\Omega}(\theta,\phi)\\
\times \, e^{ik[r\sin\theta\cos(\phi-\varphi)+z\cos\theta]}
\sin\theta\,d\phi\,d\theta.
\label{eq:rw}
\end{eqnarray}
In Eq.~(\ref{eq:rw}), the reference-sphere field $\mathbf{E}_{\Omega}$ includes the pupil field, vectorial polarization transformation, and lens apodization (see Appendix~B). For a homogeneous focusing medium of refractive index $n$ (we used the refractive index of glass, $n=1.515$, for field calculation), $\sin\theta_{\max}=\mathrm{NA}/n$ and $k=2\pi n/\lambda_0$, with $\lambda_0=532$\,nm. The prefactor $C$, pupil normalization, and magnetic-field calculation use consistent phase and medium conventions. Finally, the topology of the resulting Stokes textures is assessed through the skyrmion number defined as
\begin{equation}
N =\frac{1}{4\pi}\iint_{\mathcal D}
\mathbf{s}\cdot\left(\frac{\partial\mathbf{s}}{\partial x} \times \frac{\partial\mathbf{s}}{\partial y}\right)\,dx\,dy,
\label{eq:skyrmion-number}
\end{equation}
where $\mathcal D$ denotes the transverse domain over which the integration is performed. The boundary of $\mathcal D$ is determined using a gradient-based active-contour algorithm. Starting from a circular contour centered at the intensity centroid, the boundary is iteratively deformed toward the extremal values of the normalized Stokes parameter $s_3$ ($s_3 \simeq \pm 1$), with contour smoothing applied during the iterations. Ideally, this yields $N=\sigma$, with deviations typically up to $5\%$ when evaluated from experimental data and up to $1\%$ when evaluated from simulated data. In addition, the corresponding coverage of the Poincaré sphere provides a complementary visual assessment of the topology, as reported in Fig.~\ref{fig:Figure4}(b) for the experimental case.

\begin{figure}[!t]
\includegraphics[width=1\linewidth]{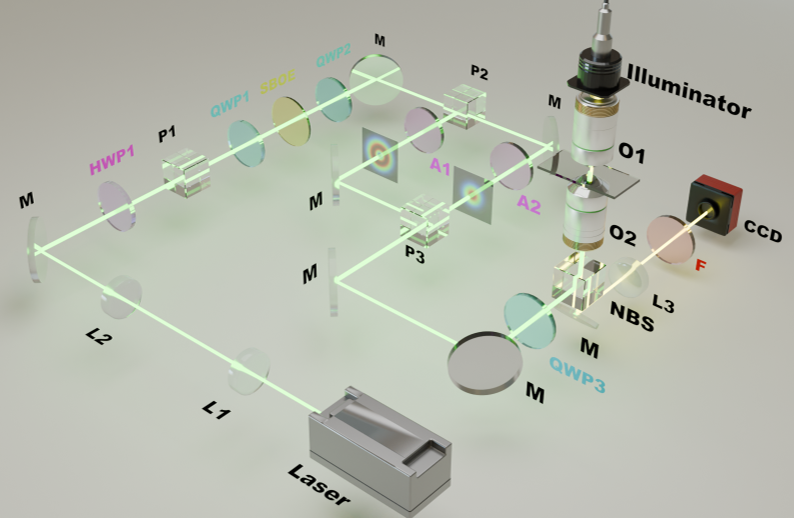}
\caption{\label{fig:Fig_setup}
Sketch of the optical trapping and manipulation experimental setup. During particle motion measurements, a filter is placed in front of the CCD camera. HWP + P: half-wave-plate and polarizer acting as an optical power control module; QWP: quarter-wave-plate; L: lens, with the pair of lenses (L1,L2) forming a telescope that resizes the collimated laser beam to a waist radius $w_{\rm in}=1.17w_{\rm design}$; P: polarizing beam splitter; NBS: non-polarizing beam splitter; M: mirror; SBOE: structured birefringence optical element (skyrmionic wave-plate); A: attenuator; O: objective; F: filter.}
\end{figure}

\subsection{Optical manipulation of trapped microparticles}

\begin{figure}[b!]
\includegraphics[width=1\linewidth]{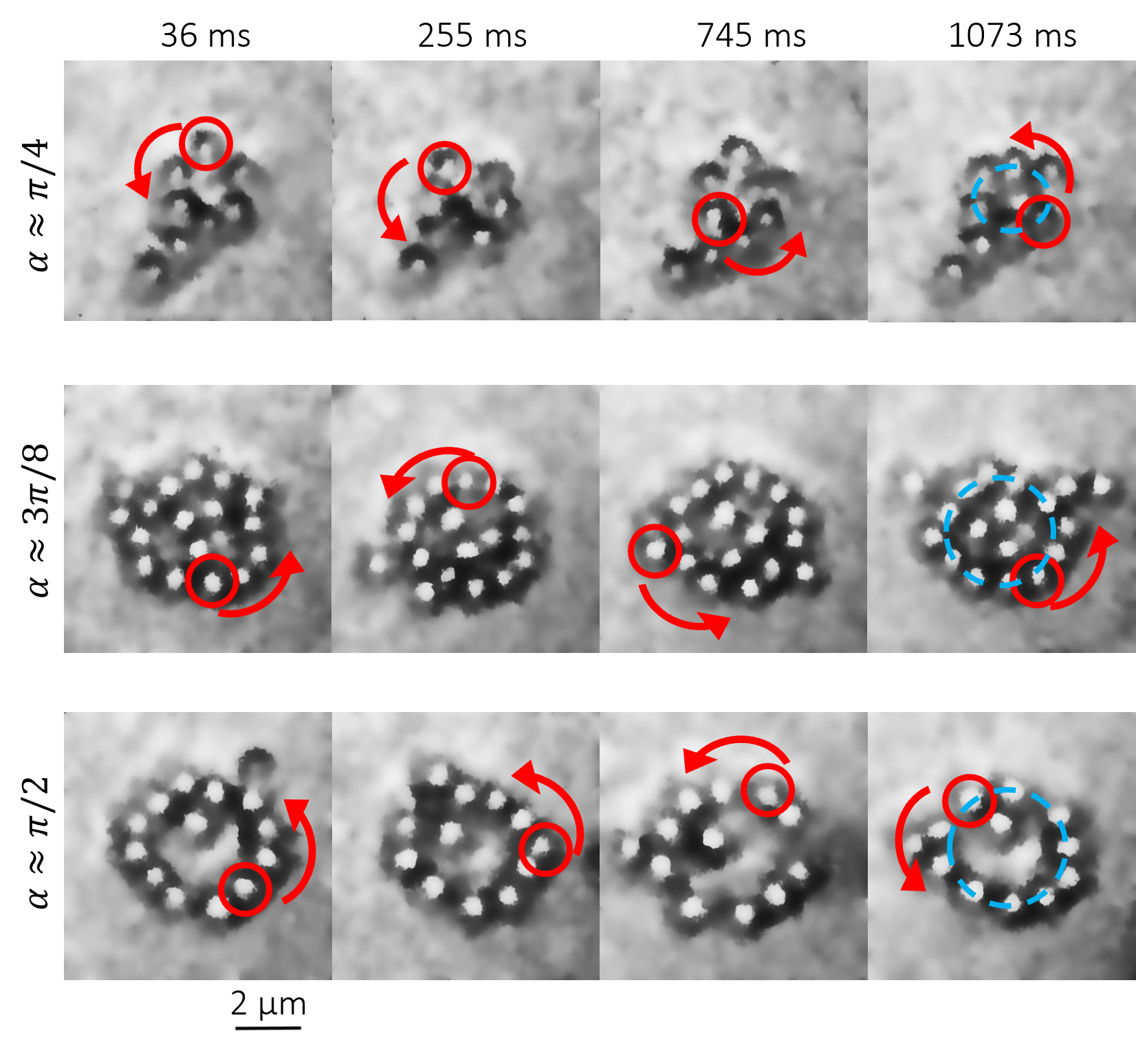}
\caption{\label{fig:Figure6}
Experimental snapshots of ring-trapped polystyrene particles for increasing values of $\alpha$. The red circle indicates the trajectory of a representative particle, while the blue dashed circle indicates its circular orbit.}
\end{figure}

The tightly focused skyrmionic beam provides sufficient optical intensity to trap and manipulate polystyrene microparticles with a diameter of approximately $1~\mu$m suspended in a water layer between two flat glass substrates. The trapping beam is focused vertically upward, thereby confining the particles along the axial direction. We consider three representative values of $\alpha$, $\alpha\approx \pi/4$, $3\pi/8$, and $\pi/2$, corresponding to equal NV/V weights, a vortex-dominated superposition, and a pure vortex beam, respectively. Due to the different transmission efficiencies of the modal weight module, the optical powers measured before the microscope objective are 46.1 mW, 37.94 mW, and 26.5 mW for these three configurations, respectively. As shown in Fig.~\ref{fig:Figure6}, orbital motion of particles confined in circular orbits is observed in all three cases, with a radius that increases with $\alpha$. The observed radial trend therefore points to a modification of the spatial beam structure rather than to a controlled variation of the optical power.

To investigate the origin of this radial trend, we first calculate the focal field using the Richards-Wolf integral within the dominant-mode approach given by Eq.~(2). For this optical comparison, the total pupil power is fixed at 100\,mW. As the vortex weight increases, the maximum of the azimuthally averaged electric-field intensity shifts from the optical axis to a radial position of about $0.19~\mu$m for the pure-vortex configuration. The mixed modes also generate angular modulations of the longitudinal field through interference. We then calculate the optical force acting on a single polystyrene sphere using finite-difference time-domain (FDTD) simulations. The complex field distribution obtained from the Richards--Wolf integral (Eq.~(\ref{eq:rw})) at a plane $1\,\mu\mathrm{m}$ before the nominal focus is imported into FDTD as the incident source. The propagation and scattering of this focused field in the presence of the particle are then simulated. The optical force is evaluated using the time-averaged Maxwell stress tensor $\langle\mathbf{T}\rangle$ according to
\begin{equation}
 \langle\mathbf F\rangle =
 \oint_{\mathcal A}\langle \mathbf{T} \rangle\cdot
       \hat{\mathbf n}\,d{\mathcal A} ,
 \label{eq:main_mst_force}
\end{equation}
where $\mathcal A$ is a closed surface surrounding the sphere and $\hat{\mathbf n}$ is its outward unit normal. The Maxwell stress tensor is given by
\begin{equation}
\begin{aligned}
\langle T_{ij}\rangle
=&\frac{1}{2}\mathrm{Re}
\left[
\epsilon E_i E_j^*
+\mu H_i H_j^*
\right]\\
&-\frac{1}{4}
\left(
\epsilon|\mathbf{E}|^2
+\mu|\mathbf{H}|^2
\right)\delta_{ij},
\end{aligned}
\end{equation}
where $\epsilon$ and $\mu$ are the permittivity and permeability of the surrounding medium, respectively. The transverse optical force distribution is obtained by scanning the particle position and repeating the FDTD calculation at each position. The simulations use refractive indices of 1.5983 for polystyrene and 1.333 for water at wavelength $\lambda_0$. The simulation domain is $5\times5\times2.2$~\textmu$\mathrm{m}^3$ with perfectly matched layers.

\begin{figure*}[t!]
\includegraphics[width=2\columnwidth]{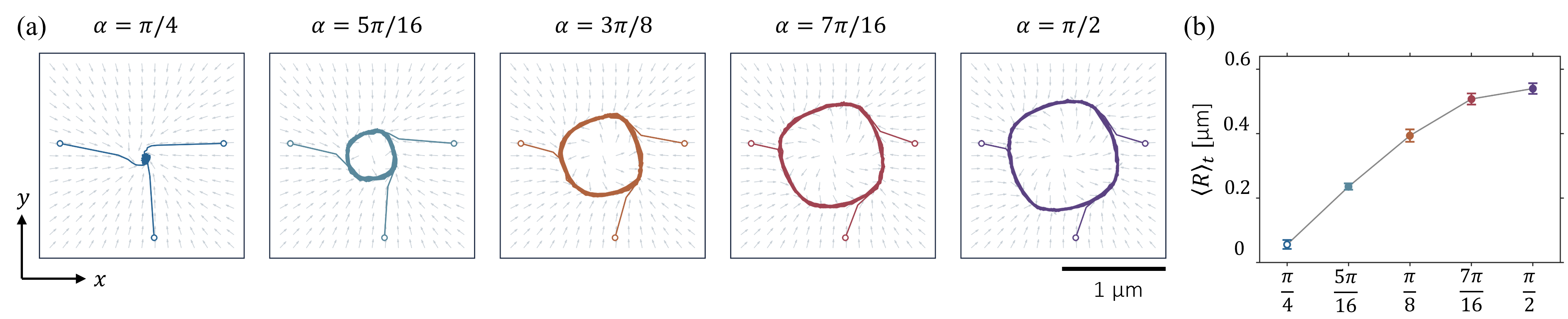}
\caption{\label{fig:Figure7}
(a) Simulated single-particle trajectories in the updated transverse force fields for different values of $\alpha$. Gray arrows indicate the transverse force distribution. Solid curves show the particle trajectories for a set of initial positions denoted by open markers. (b) Mean (markers) and standard deviation (error bars) of the particle distance from the origin in the steady state as function of $\alpha$.}
\end{figure*}

The resulting force fields reveal the evolution of the trapping potential from a near-axis confinement region toward an annular potential as the vortex weight increases. The same calculation shows a redistribution of the azimuthal energy flux \(P_\varphi=-P_x\sin\varphi+P_y\cos\varphi\), where \(\mathbf P=\tfrac12\operatorname{Re}(\mathbf E\times\mathbf H^*)\). While the potential landscape determines the radial confinement, sustained orbital motion requires work from the nonconservative component of the optical force over a complete revolution. We therefore integrate the particle dynamics using the inertial Langevin  $m\ddot{\mathbf r}+\gamma\dot{\mathbf r} =\mathbf F_{xy}+\boldsymbol{\xi}(t)$, where $\boldsymbol{\xi}$ is the thermal random contribution. Fig.~\ref{fig:Figure7}(a) shows representative trajectories obtained from simulations at 300 K while Appendix~C describes the stochastic integration and the steady-state radius statistics.
As $\alpha$ increases from the equal-weight configuration toward the pure-vortex configuration, the simulated mean orbital radius increases from a near-axis confinement at $\alpha=\pi/4$ to 0.236, 0.394, 0.507, and 0.540~\textmu m for $\alpha=5\pi/16$, $3\pi/8$, $7\pi/16$, and $\pi/2$, respectively, with within-revolution standard deviations ranging from 9.8 to 19.1 nm (Fig.~\ref{fig:Figure7}). This outward trend qualitatively agrees with the experimental observations (Fig.~\ref{fig:Figure6}). Note that, at the equal-weight configuration ($\alpha=\pi/4$), the simulation does not reproduce the experimentally observed orbital motion, as the particle remains confined near the optical axis. This discrepancy may result from the fact that the simulation considers an isolated particle at a fixed axial position, whereas the experiment involves multiple particles trapped slightly away from the nominal focal plane.

\begin{figure}[t!]
\includegraphics[width=1\columnwidth]{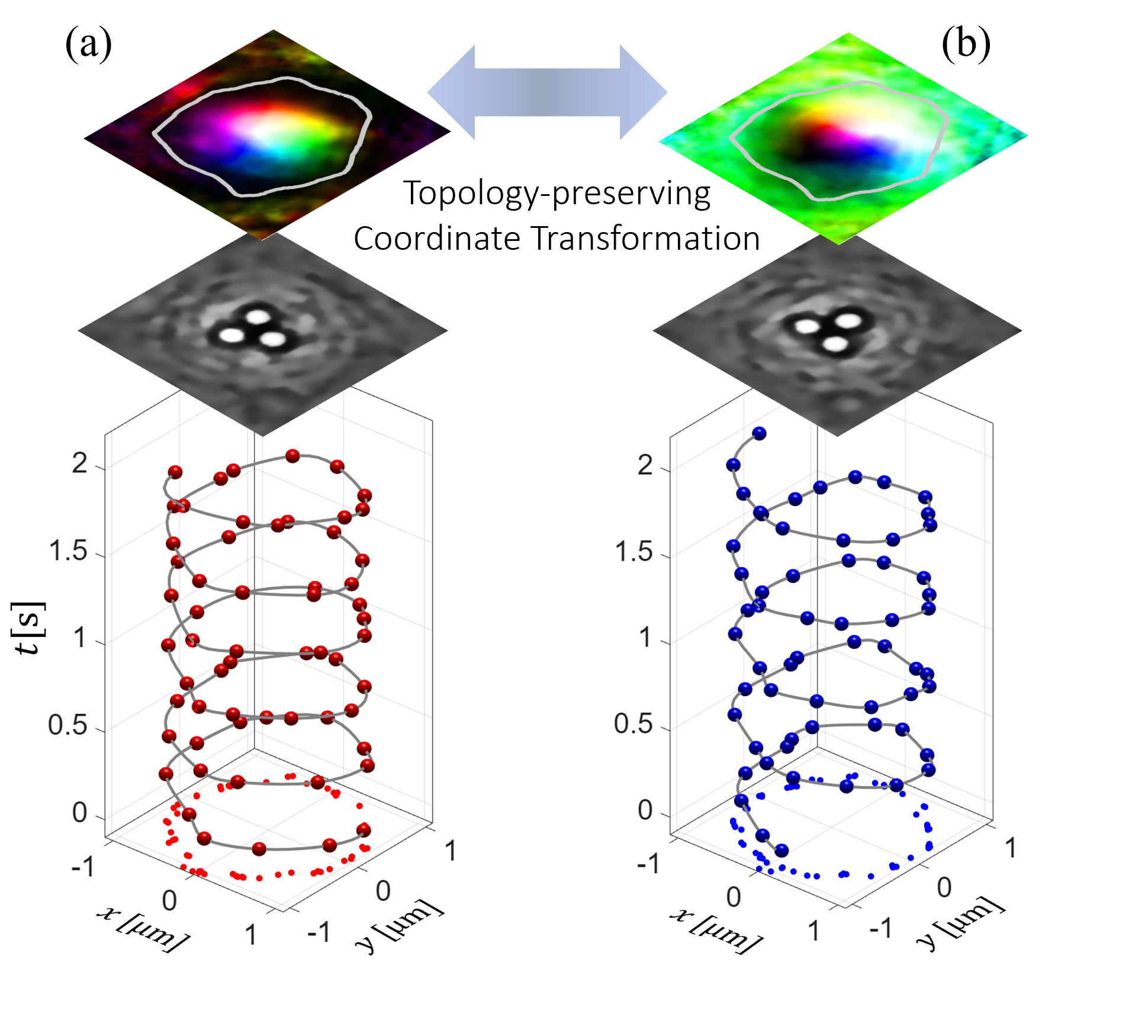}
\caption{\label{fig:Figure8} Optical trapping and manipulation of microparticles under distinct polarization textures at fixed skyrmionic topology. (a) First-order Stokes skyrmion. (b) Stokes bimeron. From top to bottom: experimental Stokes textures (see Fig.~\ref{fig:Figure4} for the color encoding); snapshot of three trapped microparticles; spatiotemporal trajectory of a given microparticle.}
\end{figure}

To further investigate the role of the polarization texture, we consider two distinct skyrmionic configurations with the same topological charge $N=1$: a unit-charge skyrmion and a bimeron. The transition between these two textures is achieved by rotating the QWP placed after the modal weight module by an angle $\pi/4$. The experiment is performed without modal weighting ($\alpha \approx \alpha_0$), with an input power close to 100\,mW, providing stable and uniform orbital rotation conditions. The results are shown in Fig.~\ref{fig:Figure8}, where the corresponding Stokes maps are also presented. 

\section{Conclusion}
\label{sec:discussion}

The present study extends optical skyrmions from the generation and characterization of topological polarization textures to the experimental manipulation of microparticles. By tightly focusing a skyrmionic beam, we experimentally demonstrate particle trapping and orbital motion using structured polarization fields. The control of the modal composition provides an experimentally accessible degree of freedom to tune particle dynamics at fixed topology, allowing the orbital radius to be modified without changing the underlying spatial modes. This modal dependence is explained by the calculated focal fields and optical forces, which reveal how conservative confinement and nonconservative driving jointly determine the emergence of orbital motion. Furthermore, the comparison between skyrmion and bimeron polarization textures with the same skyrmion number shows that distinct polarization configurations can lead to qualitatively similar optomechanical responses under the investigated conditions. While this observation suggests that the dynamics may be insensitive to the specific realization of the polarization texture, it does not by itself establish a topological invariance of the optomechanical response. More generally, these results motivate further investigation into how the structural degrees of freedom and topological properties of optical fields govern their optomechanical response.

\section*{APPENDIX A\\Laguerre-Gaussian mode decomposition}
\label{app:lg}

Within the paraxial regime, the complex field representation of the normalized Laguerre-Gaussian mode in the cylindrical coordinate system $(\rho,\phi,z)$ is given at the beam waist by (omitting the time oscillating phase term
\begin{eqnarray}
\nonumber
{\rm LG}_{l,p}^{(w_0)}(\rho,\phi,0)&=&
\sqrt{\frac{2p!}{\pi(p+|l|)!}}\frac{1}{w_0}
\left(\frac{\sqrt{2}\rho}{w_0}\right)^{|l|}\\
&\times& L_p^{|l|}\!\left(\frac{2\rho^2}{w_0^2}\right)
e^{-\rho^2/w_0^2}e^{il\phi}\,,
\label{eq:lg-waist}
\end{eqnarray}
with $L_p^{|l|}(x) = \sum_{m=0}^{p} \frac{(|l|+p)!}{(|l|+m)!(p-m)!m!}(-x)^m$. Any normalized paraxial scalar field $E$, satisfying $\iint |E|^2 \rho\,d\rho\,d\phi = 1$, can be uniquely decomposed in a given orthonormal Laguerre-Gaussian basis as
\begin{equation}
E =\sum _{l,p} c_{l,p}\,{\rm LG}_{l,p}^{(w_0)}\,,
\label{eq:decomposition}
\end{equation}
where the coefficients $c_{l,p}$ are complex numbers that depend on the choice of the basis waist radius $w_0$ and satisfy $\sum_{l,p}|c_{l,p}|^2=1$. The quantity $|c_{l,p}|^2$ thus represents the fraction of the total power carried by mode $(l,p)$ in the field $E$. The modal coefficients are obtained by projection onto the basis
\begin{equation}
c_{l,p} = \iint E \left[{\rm LG}_{l,p}^{(w_0)}\right]^*\rho d\rho d\phi \,,
\label{eq:decomposition}
\end{equation}
where the asterisk denotes complex conjugation.
\vfill

\section*{APPENDIX B\\Pupil fields for vectorial focusing}
\label{app:vectorial_focus}

In the Richards--Wolf integral, we use the normalized pupil coordinate
$q=\rho/R_p=\sin\theta/\sin\theta_{\max}$ and pupil azimuth $\phi$. With $\beta_j=R_p/w_j$, the two normalized Laguerre-Gaussian profiles in the dominant-mode calculation of Fig.~\ref{fig:Figure4}(c) are
\begin{equation}
 \begin{aligned}
 u_0(q)&=C_0e^{-(\beta_{\rm NV}q)^2},\\
 u_\sigma(q,\phi)&=
 C_1\sqrt{2}\beta_{\rm V}q\,e^{-(\beta_{\rm V}q)^2}
 e^{i\sigma\phi}.
 \end{aligned}
 \label{eq:app_pupil_mode}
\end{equation}
Here $\beta_{\rm NV}=2.61/1.76$ and $\beta_{\rm V}=\sqrt{2}\beta_{\rm NV}$. The constants $C_0$ and $C_1$ give equal basis powers over the finite pupil. We combine the two profiles with the coefficients $\cos\alpha$ and $i\sin\alpha$ in the respective circular polarizations, with $\alpha=0.30\pi$ for Fig.~\ref{fig:Figure4}(c).

For the full wave-plate calculation, $\Delta_p=2\arcsin[\sqrt{2}\beta_{\rm d}q\exp(1/2-(\beta_{\rm d}q)^2)]$. The pupil field is proportional to
\begin{equation}
 e^{-(\beta_{\rm in}q)^2-0.65i\Delta_p}
 \left[
 \cos\frac{\Delta_p}{2}\,\mathbf e_\sigma
 +i\sin\frac{\Delta_p}{2}\,e^{i\sigma\phi}\mathbf e_{-\sigma}
 \right],
 \label{eq:app_full_plate_pupil}
\end{equation}
Here $\beta_{\rm d}=R_p/w_{\rm design}$ and $\beta_{\rm in}=R_p/w_{\rm in}$. For the numerical comparison in Fig.~\ref{fig:Figure4}(c), both effective ratios are set to 2.61/1.76.

In both models, the aplanatic reference-sphere field is
\begin{equation}
 \mathbf E_\Omega=\sqrt{\cos\theta}
 \left[E_\rho(\cos\theta\,\mathbf e_\rho-\sin\theta\,\mathbf e_z)
       +E_\phi\mathbf e_\phi\right],
 \label{eq:app_reference_sphere}
\end{equation}
where $E_\rho$ and $E_\phi$ are the radial and azimuthal components of the pupil field. We use $\lambda_0=532~\mathrm{nm}$, $n=1.515$, $\mathrm{NA}=1.45$, and $f=1.8~\mathrm{mm}$ for the homogeneous-medium calculation.

\section*{APPENDIX~C\\Numerical integration of particle trajectories}
\label{app:orbit_dynamics}

For Fig.~\ref{fig:Figure7}, the transverse forces on a 1~\textmu m-diameter polystyrene sphere ($n=1.5983$) in water ($n=1.333$) were calculated by integrating the time-averaged Maxwell stress tensor at $532~\mathrm{nm}$. The FDTD domain was $5\times5\times2.2$~\textmu$\mathrm{m}^3$ with perfectly matched layers. The incident field was imported 1~\textmu m before the nominal focus plane. 

We integrate the in-plane Langevin equations at fixed axial position,
\begin{equation}
 \begin{aligned}
 d\mathbf r &= \mathbf vdt,\\
 m\,d\mathbf v =
 [\mathbf F_{xy}(\mathbf r;\alpha)-&\gamma\mathbf v]dt
 +\sqrt{2\gamma k_{\rm B}T}\,d\mathbf W_t ,
 \end{aligned}
 \label{eq:app_orbit_ode}
\end{equation}
The thermal random force $\boldsymbol{\xi}(t)$ in the main text delivers the impulse $\boldsymbol{\xi}(t)\,dt=\sqrt{2\gamma k_{\rm B}T}\,d\mathbf W_t$. The two independent Wiener increments describe the random impulses along $x$ and $y$. We take $\gamma=6\pi\eta a$ and $m=4\pi a^3\rho_{\rm p}/3$, with $a=0.50$~\textmu m, $\eta=10^{-3}~\mathrm{Pa\,s}$, $\rho_{\rm p}=1050\,\mathrm{kg\,m^{-3}}$, and $T=300~\mathrm K$. The resulting momentum relaxation time is $\tau_m=m/\gamma=5.83\times10^{-8}~\mathrm s$.

We implement a fixed-step exponential predictor/corrector in MATLAB. For a constant force over a step, it integrates the viscous velocity relaxation analytically and draws the correlated Gaussian increments in position and velocity jointly. The force is evaluated at the current and predicted positions, and their average is used for the corrected step. We use a step of 2.5~\textmu s and check it against 5~\textmu s using the same Brownian realizations for 48 trajectories per setting. For Fig.~\ref{fig:Figure7}(a), the example traces start at (0.80,0.12), (-0.80,0.12), and (0.12,-0.80), with zero initial velocity.

A fixed point is identified by convergence to a vanishing force and local linear stability. For a closed orbit, we check convergence of successive crossings of the positive $x$ axis and sample the last complete turn [$t_1,t_2]$ at 1201 equally spaced times. Writing $R(t)=\sqrt{x(t)^2+y(t)^2}$ and $T=t_2-t_1$, the statistics in Fig.~\ref{fig:Figure7}(b) are
\begin{equation}
 \begin{aligned}
 \overline R &= T^{-1}\int_{t_1}^{t_2} R(t)\,dt,\\
 \sigma_R^2 &= T^{-1}\int_{t_1}^{t_2}
        [R(t)-\overline R]^2\,dt .
 \end{aligned}
 \label{eq:app_orbit_radius}
\end{equation}
We evaluate these integrals by the trapezoidal rule for the trajectory starting at (0.80,0.12). For the equal-weight fixed point ($\alpha=\pi/4$), $\overline R$ is its distance from the optical axis and $\sigma_R=0$.

\bibliography{reference}

@article{yangOpticalSortingPresent2025,
  title = {Optical Sorting: Past, Present and Future},
  shorttitle = {Optical Sorting},
  author = {Yang, Meng and Shi, Yuzhi and Song, Qinghua and Wei, Zeyong and Dun, Xiong and Wang, Zhiming and Wang, Zhanshan and Qiu, Cheng-Wei and Zhang, Hui and Cheng, Xinbin},
  year = 2025,
  journal = {Light: Science \& Applications},
  volume = {14},
  number = {1},
  pages = {103},
  publisher = {Nature Publishing Group},
  doi = {10.1038/s41377-024-01734-5},
  copyright = {2025 The Author(s)}
}

@article{zhuMultidimensionalOpticalTweezers2023,
  title = {Multidimensional Optical Tweezers Synthetized by Rigid-Body Emulated Structured Light},
  author = {Zhu, Liuhao and Tai, Yuping and Li, Hehe and Hu, Huajie and Li, Xinzhong and Cai, Yangjian and Shen, Yijie},
  year = 2023,
  journal = {Photonics Research},
  volume = {11},
  number = {9},
  pages = {1524--1534},
  publisher = {Optica Publishing Group},
  doi = {10.1364/PRJ.490103},
  copyright = {\copyright{} 2023 Chinese Laser Press}
}

@article{gaoParaxialSkyrmionicBeams2020,
  title = {Paraxial Skyrmionic Beams},
  author = {Gao, Sijia and Speirits, Fiona C. and Castellucci, Francesco and {Franke-Arnold}, Sonja and Barnett, Stephen M. and G{\"o}tte, J{\"o}rg B.},
  year = 2020,
  journal = {Physical Review A},
  volume = {102},
  number = {5},
  pages = {053513},
  publisher = {American Physical Society},
  doi = {10.1103/PhysRevA.102.053513}
}

@inproceedings{odonnellOpticalForcesTorques2024,
  title = {Optical Forces and Torques in Skyrmionic Beams},
  booktitle = {Optical {{Trapping}} and {{Optical Micromanipulation XXI}}},
  author = {O'Donnell, Michael J. and Hanna, Simon},
  editor = {{Rubinsztein-Dunlop}, Halina and Dholakia, Kishan and Volpe, Giovanni},
  year = 2024,
  pages = {54},
  publisher = {SPIE},
  address = {San Diego, United States},
  doi = {10.1117/12.3029919},
  isbn = {978-1-5106-7884-2 978-1-5106-7885-9}
}

@article{shenFreespaceTopologicalOptical2025,
  title = {Free-Space Topological Optical Textures: Tutorial},
  shorttitle = {Free-Space Topological Optical Textures},
  author = {Shen, Yijie and Wang, Haiwen and Fan, Shanhui},
  year = 2025,
  journal = {Advances in Optics and Photonics},
  volume = {17},
  number = {2},
  pages = {295},
  doi = {10.1364/AOP.547634}
}

@article{chenGouyPhaseInduced2025,
  title = {Gouy {{Phase Induced Optical Skyrmion Transformation}} in {{Diffraction Limited Scale}}},
  author = {Chen, Jian and Shen, Xi and Zhan, Qiwen and Qiu, Cheng-Wei},
  year = 2025,
  journal = {Laser \& Photonics Reviews},
  volume = {19},
  number = {2},
  pages = {2400327},
  doi = {10.1002/lpor.202400327}
}

@article{hakobyanUnitarySpinorbitOpticalskyrmionic2024,
  title = {Unitary Spin-Orbit Optical-Skyrmionic Wave Plates},
  author = {Hakobyan, Vagharshak and Shen, Yijie and Brasselet, Etienne},
  year = 2024,
  journal = {Physical Review Applied},
  volume = {22},
  number = {5},
  pages = {054038},
  doi = {10.1103/PhysRevApplied.22.054038}
}

@article{hakobyan2024single,
  title={Single-stage spin-orbit Laguerre-Gaussian modal beam shaping from silica optics},
  author={Hakobyan, Vagharshak and Singh, Keshaan and Lei, Yuhao and Kazansky, Peter and Coursault, Delphine and Forbes, Andrew and Brasselet, Etienne},
  journal={Physical Review Applied},
  volume={21},
  number={6},
  pages={064003},
  year={2024},
  publisher={APS}
}

@article{ashkinObservationSinglebeamGradient1986,
  title = {Observation of a Single-Beam Gradient Force Optical Trap for Dielectric Particles},
  author = {Ashkin, A and Dziedzic, J M and Bjorkholm, J E and Chu, S},
  year = 1986,
  journal = {Optics Letters},
  volume = {11},
  pages = {288--290}
}

@article{volpeRoadmapOpticalTweezers2023,
  title = {Roadmap for Optical Tweezers},
  author = {Volpe, Giovanni and Marag{\`o}, Onofrio M and {Rubinsztein-Dunlop}, Halina and Pesce, Giuseppe and Stilgoe, Alexander B and Volpe, Giorgio and Tkachenko, Georgiy and Truong, Viet Giang and Chormaic, S{\'i}le Nic and Kalantarifard, Fatemeh and Elahi, Parviz and K{\"a}ll, Mikael and Callegari, Agnese and Marqu{\'e}s, Manuel I and Neves, Antonio A R and Moreira, Wendel L and Fontes, Adriana and Cesar, Carlos L and Saija, Rosalba and Saidi, Abir and Beck, Paul and Eismann, J{\"o}rg S and Banzer, Peter and Fernandes, Thales F D and Pedaci, Francesco and Bowen, Warwick P and Vaippully, Rahul and Lokesh, Muruga and Roy, Basudev and {Thalhammer-Thurner}, Gregor and {Ritsch-Marte}, Monika and Garc{\'i}a, Laura P{\'e}rez and Arzola, Alejandro V and Castillo, Isaac P{\'e}rez and Argun, Aykut and Muenker, Till M and Vos, Bart E and Betz, Timo and Cristiani, Ilaria and Minzioni, Paolo and Reece, Peter J and Wang, Fan and McGloin, David and Ndukaife, Justus C and Quidant, Romain and Roberts, Reece P and Laplane, Cyril and Volz, Thomas and Gordon, Reuven and Hanstorp, Dag and Marmolejo, Javier Tello and Bruce, Graham D and Dholakia, Kishan and Li, Tongcang and Brzobohat{\'y}, Oto and Simpson, Stephen H and Zem{\'a}nek, Pavel and Ritort, Felix and Roichman, Yael and Bobkova, Valeriia and Wittkowski, Raphael and Denz, Cornelia and Kumar, G V Pavan and Foti, Antonino and Donato, Maria Grazia and Gucciardi, Pietro G and Gardini, Lucia and Bianchi, Giulio and Kashchuk, Anatolii V and Capitanio, Marco and Paterson, Lynn and Jones, Philip H and {Berg-S{\o}rensen}, Kirstine and Barooji, Younes F and Oddershede, Lene B and Pouladian, Pegah and Preece, Daryl and Adiels, Caroline Beck and De Luca, Anna Chiara and Magazz{\`u}, Alessandro and Bronte Ciriza, David and Iat{\`i}, Maria Antonia and Swartzlander, Grover A},
  year = 2023,
  journal = {Journal of Physics: Photonics},
  volume = {5},
  number = {2},
  pages = {022501},
  publisher = {IOP Publishing},
  doi = {10.1088/2515-7647/acb57b}
}

@article{yangOpticalTrappingStructured2021,
  title = {Optical Trapping with Structured Light: A Review},
  shorttitle = {Optical Trapping with Structured Light},
  author = {Yang, Yuanjie and Ren, Yuxuan and Chen, Mingzhou and Arita, Yoshihiko and {Rosales-Guzm{\'a}n}, Carmelo},
  year = 2021,
  journal = {Advanced Photonics},
  volume = {3},
  number = {3},
  pages = {034001},
  publisher = {SPIE},
  doi = {10.1117/1.AP.3.3.034001}
}

@article{otteOpticalTrappingGets2020,
  title = {Optical Trapping Gets Structure: {{Structured}} Light for Advanced Optical Manipulation},
  shorttitle = {Optical Trapping Gets Structure},
  author = {Otte, E. and Denz, C.},
  year = 2020,
  journal = {Applied Physics Reviews},
  volume = {7},
  number = {4},
  pages = {041308},
  doi = {10.1063/5.0013276}
}

@article{zhuOpticalSkippingRope2023,
  title = {Optical Skipping Rope Induced Transverse {{OAM}} for Particle Orbital Motion Parallel to the Optical Axis},
  author = {Zhu, Liuhao and Zhang, Xiaohe and Rui, Guanghao and He, Jun and Gu, Bing and Zhan, Qiwen},
  year = 2023,
  journal = {Nanophotonics},
  volume = {12},
  number = {23},
  pages = {4351--4359},
  doi = {10.1515/nanoph-2023-0551}
}

@article{gaoParticleDeliveryGeneralized2024a,
  title = {Particle Delivery in Generalized Optical Vortex Conveyor Belts with a Uniform Orbital Flow},
  author = {Gao, Wenyu and Zhou, Yuan and Li, Xing and Zhang, Yanan and Zhang, Qiang and Li, Manman and Yu, Xianghua and Yan, Shaohui and Xu, Xiaohao and Yao, Baoli},
  year = 2024,
  journal = {Photonics Research},
  volume = {12},
  number = {12},
  pages = {2881--2890},
  publisher = {Optica Publishing Group},
  doi = {10.1364/PRJ.539718},
  copyright = {\copyright{} 2024 Chinese Laser Press}
}

@article{shenOpticalVortices302019,
  title = {Optical Vortices 30 Years on: {{OAM}} Manipulation from Topological Charge to Multiple Singularities},
  shorttitle = {Optical Vortices 30 Years On},
  author = {Shen, Yijie and Wang, Xuejiao and Xie, Zhenwei and Min, Changjun and Fu, Xing and Liu, Qiang and Gong, Mali and Yuan, Xiaocong},
  year = 2019,
  journal = {Light: Science \& Applications},
  volume = {8},
  number = {1},
  pages = {90},
  publisher = {Nature Publishing Group},
  doi = {10.1038/s41377-019-0194-2},
  copyright = {2019 The Author(s)}
}

@article{gaoTopologicalLightField2025,
  title = {Topological Light Field: From Singular to Skyrmionic Optics and Beyond},
  shorttitle = {Topological Light Field},
  author = {Gao, Wenyu and Zhou, Yuan and Li, Xing and Zhang, Yanan and Zhang, Qiang and Li, Manman and Yu, Xianghua and Yan, Shaohui and Xu, Xiaohao and Yao, Baoli},
  year = 2025,
  journal = {Journal of Optics},
  volume = {27},
  number = {8},
  pages = {083001},
  publisher = {IOP Publishing},
  doi = {10.1088/2040-8986/adf3c4}
}

@article{shenGenerationOpticalSkyrmions2022,
  title = {Generation of {{Optical Skyrmions}} with {{Tunable Topological Textures}}},
  author = {Shen, Yijie and Mart{\'i}nez, Eduardo Casas and {Rosales-Guzm{\'a}n}, Carmelo},
  year = 2022,
  journal = {ACS Photonics},
  volume = {9},
  number = {1},
  pages = {296--303},
  doi = {10.1021/acsphotonics.1c01703}
}

@article{tsessesOpticalSkyrmionLattice2018,
  title = {Optical Skyrmion Lattice in Evanescent Electromagnetic Fields},
  author = {Tsesses, S. and Ostrovsky, E. and Cohen, K. and Gjonaj, B. and Lindner, N. H. and Bartal, G.},
  year = 2018,
  journal = {Science},
  volume = {361},
  number = {6406},
  pages = {993--996},
  publisher = {American Association for the Advancement of Science},
  doi = {10.1126/science.aau0227}
}

@article{dengObservationLocalizedMagnetic2022,
  title = {Observation of Localized Magnetic Plasmon Skyrmions},
  author = {Deng, Zi-Lan and Shi, Tan and Krasnok, Alex and Li, Xiangping and Al{\`u}, Andrea},
  year = 2022,
  journal = {Nature Communications},
  volume = {13},
  number = {1},
  pages = {8},
  publisher = {Nature Publishing Group},
  doi = {10.1038/s41467-021-27710-w},
  copyright = {2022 The Author(s)}
}

@article{guoSelfhealingOpticalSkyrmionic2025,
  title = {Self-Healing of Optical Skyrmionic Beams},
  author = {Guo, Haochen and Das, Trishita and Wu, Haijun and Dev, Vasu and Zhu, Zhihan and Shen, Yijie},
  year = 2025,
  journal = {Journal of Optics},
  volume = {27},
  number = {2},
  pages = {025604},
  publisher = {IOP Publishing},
  doi = {10.1088/2040-8986/adab84}
}

@article{duDeepsubwavelengthFeaturesPhotonic2019,
  title = {Deep-Subwavelength Features of Photonic Skyrmions in a Confined Electromagnetic Field with Orbital Angular Momentum},
  author = {Du, Luping and Yang, Aiping and Zayats, Anatoly V. and Yuan, Xiaocong},
  year = 2019,
  journal = {Nature Physics},
  volume = {15},
  number = {7},
  pages = {650--654},
  publisher = {Nature Publishing Group},
  doi = {10.1038/s41567-019-0487-7},
  copyright = {2019 The Author(s), under exclusive licence to Springer Nature Limited}
}

@article{wuPhotonicTorons3D2025,
  title = {Photonic {{Torons}} with {{3D Topology Transitions}} and {{Tunable Spin Monopoles}}},
  author = {Wu, Haijun and {Mata-Cervera}, Nilo and Wang, Haiwen and Zhu, Zhihan and Qiu, Chengwei and Shen, Yijie},
  year = 2025,
  journal = {Physical Review Letters},
  volume = {135},
  number = {6},
  pages = {063802},
  doi = {10.1103/8pb4-zy7c}
}

@article{hakobyanPlatesOpticalVortices2025,
  title = {Q -{{Plates}}: {{From Optical Vortices}} to {{Optical Skyrmions}}},
  shorttitle = {Q -{{Plates}}},
  author = {Hakobyan, Vagharshak and Brasselet, Etienne},
  year = 2025,
  journal = {Physical Review Letters},
  volume = {134},
  number = {8},
  pages = {083802},
  doi = {10.1103/PhysRevLett.134.083802}
}

@article{xieTopologicalLightWaves2025,
  title = {Topological Light Waves Manipulating Particles: {{A}} Perspective},
  shorttitle = {Topological Light Waves Manipulating Particles},
  author = {Xie, Xi and Shen, Yijie},
  year = 2025,
  journal = {Light Manipulation and Applications},
  volume = {1},
  number = {1},
  pages = {202501--202501},
  publisher = {Science Exploration Press},
  doi = {10.70401/lma.2026.0001}
}

@article{wangTopologicalWaterwaveStructures2025,
  title = {Topological Water-Wave Structures Manipulating Particles},
  author = {Wang, Bo and Che, Zhiyuan and Cheng, Cheng and Tong, Caili and Shi, Lei and Shen, Yijie and Bliokh, Konstantin Y. and Zi, Jian},
  year = 2025,
  journal = {Nature},
  volume = {638},
  number = {8050},
  pages = {394--400},
  publisher = {Nature Publishing Group},
  doi = {10.1038/s41586-024-08384-y},
  copyright = {2025 The Author(s), under exclusive licence to Springer Nature Limited}
}

@article{smirnovaWaterWaveVorticesSkyrmions2024a,
  title = {Water-{{Wave Vortices}} and {{Skyrmions}}},
  author = {Smirnova, Daria A. and Nori, Franco and Bliokh, Konstantin Y.},
  year = 2024,
  journal = {Physical Review Letters},
  volume = {132},
  number = {5},
  pages = {054003},
  publisher = {American Physical Society},
  doi = {10.1103/PhysRevLett.132.054003}
}

@article{xuTopologicalMagneticLattices2025,
  title = {Topological {{Magnetic Lattices}} for {{On-Chip Nanoparticle Trapping}} and {{Sorting}}},
  author = {Xu, Hongyang and Xie, Xi and Zhang, Chuangye and Zhang, Yuquan and Yuan, Xiaocong and Shen, Yijie and Min, Changjun},
  year = 2025,
  journal = {Nano Letters},
  volume = {25},
  number = {26},
  pages = {10611--10618},
  publisher = {American Chemical Society},
  doi = {10.1021/acs.nanolett.5c02383}
}

@article{tangAchiralChiralOptical2023,
  title = {Achiral and {{Chiral Optical Force Within Topological Optical Lattices Generated}} with {{Plasmonic Metasurfaces}} and {{Tunable Incident Beam}}},
  author = {Tang, Xi and Kuai, Yan and Fan, Zetao and Lu, Fengya and Zang, Haofeng and Chen, Junxue and Zhan, Qiwen and Zhang, Douguo},
  year = 2023,
  journal = {Physical Review Applied},
  volume = {19},
  number = {5},
  pages = {054016},
  publisher = {American Physical Society},
  doi = {10.1103/PhysRevApplied.19.054016}
}

@article{curtisDynamicHolographicOptical2002,
  author = {Curtis, Jennifer E. and Koss, Brian A. and Grier, David G.},
  title = {{Dynamic holographic optical tweezers}},
  journal = {Optics Communications},
  year = {2002},
  volume = {207},
  number = {1-6},
  pages = {169--175},
  doi = {10.1016/S0030-4018(02)01524-9},
  url = {https://doi.org/10.1016/S0030-4018(02)01524-9}
}

@article{padgettTweezersTwist2011,
  author = {Padgett, Miles and Bowman, Richard},
  title = {{Tweezers with a twist}},
  journal = {Nature Photonics},
  year = {2011},
  volume = {5},
  number = {6},
  pages = {343--348},
  doi = {10.1038/nphoton.2011.81},
  url = {https://doi.org/10.1038/nphoton.2011.81}
}

@article{rodrigoFreestyle3DLaser2015,
  author = {Rodrigo, Jos{\'e} A. and Alieva, Tatiana},
  title = {{Freestyle 3D laser traps: tools for studying light-driven particle dynamics and beyond}},
  journal = {Optica},
  year = {2015},
  volume = {2},
  number = {9},
  pages = {812--815},
  doi = {10.1364/OPTICA.2.000812},
  url = {https://doi.org/10.1364/OPTICA.2.000812}
}

@article{wuTimeVarying3DOptical2025,
  author = {Wu, Yi-Jing and Zhuang, Jing-Han and Yu, Pan-Pan and Liu, Yi-Fan and Wang, Zi-Qiang and Li, Yin-Mei and Qiu, Cheng-Wei and Gong, Lei},
  title = {{Time-varying 3D optical torque via a single beam}},
  journal = {Nature Communications},
  year = {2025},
  volume = {16},
  number = {1},
  pages = {593},
  doi = {10.1038/s41467-024-55781-y},
  url = {https://doi.org/10.1038/s41467-024-55781-y}
}

@article{yuanSpinEngineered3DOptical2025,
  author = {Yuan, Zheng and Gao, Yuan and Yan, Wenxiang and Qing, Zhiming and Ren, Zhi-Cheng and Wang, Xi-Lin and Ding, Jianping and Wang, Hui-Tian},
  title = {{Spin-engineered 3D optical manipulation via polarization-gradient architecturing}},
  journal = {Optica},
  year = {2025},
  volume = {12},
  number = {6},
  pages = {853--859},
  doi = {10.1364/OPTICA.561034},
  url = {https://doi.org/10.1364/OPTICA.561034}
}

@article{yanStructuringLightFlows2026,
  author = {Yan, Wenxiang and Yuan, Zheng and Gao, Yuan and Chen, Zhaozhong and Ren, Zhi-Cheng and Wang, Xi-Lin and Ding, Jianping and Wang, Hui-Tian},
  title = {{Structuring light with flows}},
  journal = {Nature Communications},
  year = {2026},
  volume = {17},
  number = {1},
  pages = {1817},
  doi = {10.1038/s41467-026-69117-5},
  url = {https://doi.org/10.1038/s41467-026-69117-5}
}

@article{yuanLongitudinallyTunedOAM2026,
  author = {Yuan, Zheng and Yan, Wenxiang and Ren, Zhi-Cheng and Wang, Xi-Lin and Ding, JianPing and Wang, Hui-Tian},
  title = {{Longitudinally tuned OAM vortex beams for advanced particle manipulation}},
  journal = {Optica},
  year = {2026},
  volume = {13},
  number = {4},
  pages = {651--660},
  doi = {10.1364/OPTICA.580854},
  url = {https://doi.org/10.1364/OPTICA.580854}
}

@article{zhangStructurelessExcitationManipulation2026,
  author = {Zhang, Yuquan and Ma, Haixiang and Ju, Zhendong and Chen, Xusheng and Chen, Yixuan and Xie, Xi and Du, Luping and Min, Changjun and Yuan, Xiaocong},
  title = {{Structureless excitation and manipulation of dynamic holographic plasmonic slides}},
  journal = {Nature Communications},
  year = {2026},
  volume = {17},
  number = {1},
  pages = {2946},
  doi = {10.1038/s41467-026-69879-y},
  url = {https://doi.org/10.1038/s41467-026-69879-y}
}

@article{liFreeformOpticalFlow2026,
  author = {Li, Tianyue and Li, Xiao and Gao, Zengyang and Ng, Jack and Blahnik, Vladan and Nan, Fan and Zheng, Yuebing and Chan, C. T.},
  title = {{Freeform optical flow based on meta-conveyors for compact, programmable in situ nanomanipulation}},
  journal = {Nature Communications},
  year = {2026},
  volume = {17},
  number = {1},
  pages = {4212},
  doi = {10.1038/s41467-026-73024-0},
  url = {https://doi.org/10.1038/s41467-026-73024-0}
}

@article{liScalableGeneralizedMetaSpanners2026,
  author = {Li, Tianyue and Gao, Wenyu and Fu, Boyan and Shao, Tianhua and Fu, Yuchao and Zavatski, Siarhei and Nayak, Jeeban Kumar and Yan, Shaohui and Xu, Xiaohao and Wang, Shuming and Yao, Baoli and Wang, Zhenlin and Zhu, Shining and Martin, Olivier J. F. and Chan, C. T.},
  title = {{Scalable generalized meta-spanners enabling parallel multitasking optical manipulation}},
  journal = {Science Advances},
  year = {2026},
  volume = {12},
  number = {15},
  pages = {eaed5302},
  doi = {10.1126/sciadv.aed5302},
  url = {https://doi.org/10.1126/sciadv.aed5302}
}

@article{hanDynamicCharacteristicsSkyrmionic2026,
  author  = {Han, Guoxia and Tian, Yihan and Hu, Ke and
             Zhang, Feiyang and Jing, Maoda and Wang, Guangyi and
             Zhao, Qihui and Zhao, Li},
  title   = {The dynamic characteristics of non-diffracting
             skyrmionic beams under tight focusing configurations},
  journal = {Journal of Quantitative Spectroscopy and Radiative Transfer},
  volume  = {350},
  pages   = {109756},
  year    = {2026},
  doi     = {10.1016/j.jqsrt.2025.109756}
}

@article{duanQuasicrystalTopologicalHydrodynamics2026,
  author  = {Duan, Jun-liang and Hu, Chuanjie and Ning, Jiayun and
             Chai, Xu-jie and Wang, Li-Wei and Dong, Yang and
             Liu, Jianjun and Zhu, Shan and Chen, Huanyang and
             Jiang, Jian-Hua and Chen, Jin-hui},
  title   = {Quasicrystal Topological Hydrodynamics},
  journal = {Physical Review X},
  volume  = {16},
  pages   = {031078},
  year    = {2026},
  doi     = {10.1103/rfgr-n77g}
}

@misc{zhangSkyrmionMicroparticleSAW2026,
  author        = {Zhang, Jiaqi and Wu, Decai and Ma, Tingfeng and Lou, Chenbowen and Wu, Bowei and Li, Shuanghuizhi},
  title = {Topological Skyrmion-type microparticle manipulation \\ based on surface acoustic wave phase modulations},
  year          = {2026},
  eprint        = {arXiv:2608.16258},
  archivePrefix = {arXiv},
  primaryClass  = {physics.app-ph},
  note          = {Preprint},
  doi           = {10.48550/arXiv.2608.16258}
}

@article{chavilkkadanPurelyGeometricSpin2025,
  title = {Purely Geometric Spin--Orbit {{Laguerre}}--{{Gauss}} Waveplates from {{3D}} Laser-Nanostructured Silica Glass},
  author = {Chavilkkadan, Mufeeduzaman and Shevtsov, Sergei and Kazansky, Peter and Brasselet, Etienne},
  year = 2025,
  journal = {APL Photonics},
  volume = {10},
  number = {4},
  pages = {040806},
  doi = {10.1063/5.0255787}
}

@article{maObservationSuperoscillationMoire2026,
  title = {Observation of Superoscillation Moir\'e Superlattices},
  author = {Ma, Xin and Zhang, Hao and Wei, Wenjun and Tai, Yuping and Qian, Yixian and Li, Xinzhong and Shen, Yijie},
  year = 2026,
  journal = {Advanced Photonics},
  volume = {8},
  number = {2},
  pages = {026008},
  publisher = {SPIE},
  doi = {10.1117/1.AP.8.2.026008}
}

@article{luNanoparticleDeepSubwavelengthDynamics2024,
  title = {Nanoparticle {{Deep-Subwavelength Dynamics Empowered}} by {{Optical Meron}}--{{Antimeron Topology}}},
  author = {Lu, Chengfeng and Wang, Bo and Fang, Xiang and Tsai, Din Ping and Zhu, Weiming and Song, Qinghua and Deng, Xiao and He, Tao and Gong, Xiaoyun and Luo, Hong and Wang, Zhanshan and Dai, Xinhua and Shi, Yuzhi and Cheng, Xinbin},
  year = 2024,
  journal = {Nano Letters},
  volume = {24},
  number = {1},
  pages = {104--113},
  doi = {10.1021/acs.nanolett.3c03351}
}

@article{wangOpticalTrappingMesoscale2025,
  title = {Optical Trapping of Mesoscale Particles and Atoms in Hollow-Core Optical Fibers: Principle and Applications},
  shorttitle = {Optical Trapping of Mesoscale Particles and Atoms in Hollow-Core Optical Fibers},
  author = {Wang, Rui and Li, Wei and Xia, Zhiwen and Deng, Hongchang and Zhang, Yao and Fu, Rongxin and Zhang, Shuailong and Euser, Tijmen G. and Yuan, Libo and Song, Ningfang and Jiang, Yi and Xie, Shangran},
  year = 2025,
  journal = {Light: Science \& Applications},
  volume = {14},
  number = {1},
  pages = {146},
  publisher = {Nature Publishing Group},
  doi = {10.1038/s41377-025-01801-5},
  copyright = {2025 The Author(s)}
}

@article{zhangPlasmonicTweezersNanoscale2021,
  title = {Plasmonic Tweezers: For Nanoscale Optical Trapping and Beyond},
  shorttitle = {Plasmonic Tweezers},
  author = {Zhang, Yuquan and Min, Changjun and Dou, Xiujie and Wang, Xianyou and Urbach, Hendrik Paul and Somekh, Michael G. and Yuan, Xiaocong},
  year = 2021,
  journal = {Light: Science \& Applications},
  volume = {10},
  number = {1},
  pages = {59},
  publisher = {Nature Publishing Group},
  doi = {10.1038/s41377-021-00474-0},
  copyright = {2021 The Author(s)}
}
\end{document}